\documentclass[aps,pra,english,twocolumn,showpacs,preprintnumbers,amsmath,amssymb,floatfix,superscriptaddress]{revtex4-1}

\usepackage[T1]{fontenc}
\usepackage[latin9]{inputenc}
\usepackage{graphicx}
\usepackage{amssymb}
\usepackage{babel}


\usepackage{babel}
\makeatother

\begin{document}

\title{Enlarged Molecules from Excited Atoms in Nanochannels}

\author{Mathias Bostr{\"o}m}
\email{mabos@ifm.liu.se}
\affiliation{Department of Applied Mathematics, Australian National University, Canberra, Australia}
\affiliation{Department of Energy and Process Engineering, Norwegian University of Science and Technology, N-7491 Trondheim, Norway}
\author{Iver Brevik}
\affiliation{Department of Energy and Process Engineering, Norwegian University of Science and Technology, N-7491 Trondheim, Norway}
\author{Bo E. Sernelius}
\email{bos@ifm.liu.se}
\affiliation{Division of Theory and Modeling, Department of Physics, 
Chemistry and Biology, Link\"{o}ping University, SE-581 83 Link\"{o}ping, Sweden}
\author{Maofeng Dou }
\affiliation{ Dept of Materials Science and Engineering, Royal Institute of Technology, SE-100 44 Stockholm, Sweden}
\author{Clas Persson}
\affiliation{ Dept of Materials Science and Engineering, Royal Institute of Technology, SE-100 44 Stockholm, Sweden}
\affiliation{Department of Physics, University of Oslo, P. Box 1048 Blindern,
NO-0316 Oslo, Norway}
\author{Barry W. Ninham}
\affiliation{Department of Applied Mathematics, Australian National University, Canberra, Australia}

\begin{abstract}
 The resonance interaction that takes  place in planar nanochannels between pairs of excited state atoms is explored. We consider interactions in channels of silica, zinc oxide and gold.  The nanosized channels induce a dramatically  different interaction from that in free space.  Illustrative calculations for two lithium and cesium atoms, demonstrate that there is a short range repulsion followed by long range attraction. The binding energy is strongest near the surfaces. The size of the enlarged molecule is biggest at the center of the cavity and increases with channel width.  Since the interaction is generic, we predict that enlarged molecules are formed in porous structures, and that the molecule size depends on the size of the nanochannels.

\end{abstract}


\maketitle

How molecular interactions and energy transfer are affected by the shape and size of cavities and modulated by the surfaces that confine them is of interest in catalysis. Little work has been done in this area. Zeolites, mesoporous silicas, and their use for oil cracking may be one of the largest of all industries. Despite this, there is still no general agreement on the mechanisms of adsorption of enzymes and other molecules in mesoporous silica, and their excited state interactions.\,\cite{Blum}. An investigation of how molecular interactions and energy transfer are affected by the shape and size of cavities  and of the surfaces  that confine them is then of interest. Hopmeier et al.\,\cite{Hopmeier} have demonstrated experimental evidence for enhancement of dipole-dipole interaction in a microcavity and Agarwal and Gupta\,\cite{Agar} have demonstrated this in theory. An inhibition to such work  has been that  the standard theoretical expression for the resonance interaction between excited state-ground state atoms is incorrect\,\cite{Bostrom1,BostromEPL,Hartman,Sherkunov}. 

In this work we demonstrate how resonance interactions between excited atoms are strongly modified at nanoscale dimensions when the atoms interact inside planar channels. We show that the containment effects on the interaction can lead to the formation of peculiar enlarged molecules. As compared to our previous work\,\cite{BostromEPL} the present contains a deeper analysis of the phenomenon. This includes an account for the origin of the short-range repulsive and  long-range attractive interaction via spectral plots of interactions from different excitation branches and detailed studies of the effects due to different locations of the atomic species and different cavity sizes. The binding energy is strongest near the surfaces. The size of the enlarged molecule is biggest in the center of the cavity. We use lithium and cesium atoms and channels in silica, zinc oxide and gold as examples. We first briefly rehearse the (correct) theory of the resonance interaction energy in channels and in free space.  With that established we present some illustrative results. We compare the very different interactions of atoms in free space and in nanochannels. 

We have shown previously\,\cite{Bostrom1,BostromEPL} that, due to too drastic approximations, the underlying theory of resonance interactions derived from perturbative quantum electrodynamics (QED) is only correct in the non-retarded limit.
To see this we recall the standard argument:  Consider two identical atoms where one initially is in its ground state and the other is in an excited state.  This state can also be represented by a superposition of states: one symmetric and one antisymmetric with respect to interchange of the atoms. While the symmetric state is likely to decay into two ground-state atoms, the antisymmetric state can be quite long-lived. The system can thus be trapped in the antisymmetric state\,\cite{Bostrom1,Stephen}. The energy migrates back and forth between the two atoms until either the two atoms move apart or a photon is emitted away from the system. First order dispersion interactions are caused by this coupling of the system, i.e. the energy difference between the two states is separation ($\rho$) dependent. After writing down the equations of motion for the excited system it is straightforward to derive the zero temperature Green function for two identical and isotropic atoms\,\cite{McLachlan,Bostrom1,BostromEPL}. The  resonance frequencies ($\omega_r$) of the system are given by the following equation\,\cite{McLachlan}:
\begin{equation}
1-\alpha(1|\omega) \alpha(2|\omega) T(\rho|\omega)^2=0,
\label{Eq3}
\end{equation} 
where the atom polarizability is taken to have the form (for lithium and cesium)
\begin{equation}
\alpha (j|\omega ) \approx \frac{{\alpha (j|0)}}{{\omega _j^2 - {\omega ^2} - i\omega {\gamma _j}}}.
\label{Eq3b}
\end{equation}
The parameters for Li and Cs were obtained\,\cite{Boser} through a fit of the van der Waals and Casimir asymptotes to the results from a full ab initio quantum mechanical calculation by Marinescu and You\,\cite{Marinescu}. The fitting procedure for this single oscillator model works very well for all combinations of alkali-metal dimers in vacuum. For dimers immersed in a medium one may expect that a more realistic form of the polarizabilities are needed.

\begin{figure}
\includegraphics[width=8cm]{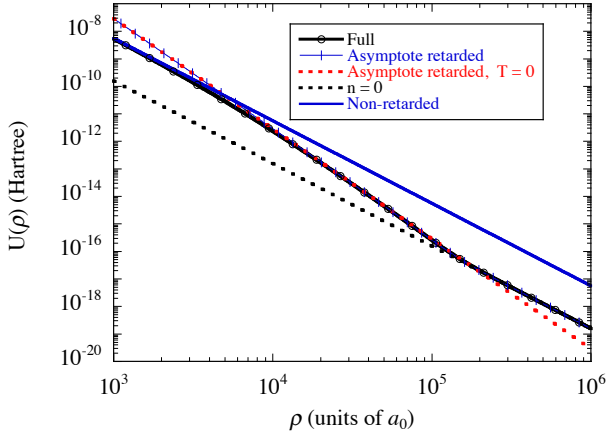}
\caption{(Color online) Full interaction energy between two lithium atoms (when the modes in the ($\pm$;$x$) branch are excited) and various limiting results as functions of separation, $\rho$. The full result from Eq.\,(\ref{Eq35}) is shown as the thick solid curve with circles; the large separation asymptote from Eq.\,(\ref{Eq35b}) is represented by a thin curve with vertical bars; the T = 0 K version of this asymptote from Eq.\,(\ref{AEq32}) is represented by the steepest of the two dotted straight lines; the other dotted line is the $n = 0$ term from Eq.\,(\ref{Eq36}); the non-retarded full result is given by the solid straight line. All results are in atomic units, i.e., the energies in Hartree units and the separation in Bohr radii. }
\label{figu1}
\end{figure}

In a vacuum the field susceptibility matrix, ${\bf  T}(\rho|i \omega)$, has the following non-zero matrix elements 
\cite{BostromEPL},
\begin{equation}
\begin{array}{*{20}{l}}
{T_{xx}^0(i\omega ) = 2(\frac{1}{{{\rho ^2}}} + \frac{{\omega }}{{\rho c}})\frac{{{e^{ - \omega \rho  /c}}}}{{\rho }},}\\
{T_{yy}^0(i\omega ) = T_{zz}^0(i\omega ) =  - (\frac{{{\omega ^2}}}{{{c^2}}} + \frac{{\omega  }}{{\rho c}} + \frac{1}{{{\rho ^2}}})\frac{{{e^{ - \omega \rho /c}}}}{{\rho}}},
\end{array}
\label{Eq4}
\end{equation}
when the atoms are aligned along the $x$-axis. 
The susceptibilities in a cavity with dielectric walls were used by Rahmani et al.\,\cite{Rahmani} to discuss fluorescence lifetimes. If the two molecules are at the distance $z$ from one of the channel boundaries the corrections to the free space results when the atoms are symmetrically excited involve the change in the trace of the matrix,  $T_{xx}^1 + T_{yy}^1 + T_{zz}^1$, where
\begin{equation}
\begin{array}{*{20}{l}}
{\begin{array}{*{20}{l}}
{T_{xx}^1 + T_{yy}^1 =  - \int_0^\infty  {\frac{{dqq{J_0}(q\rho )}}{{{\gamma _0}}}} \left\{ {\left[ {\frac{{\gamma _0^2}}{{1 - \Delta _p^2{e^{ - 2{\gamma _0}d}}}}} \right]\left( {{\Delta _p}{e^{ - 2{\gamma _0}z}}} \right.} \right.}\\
{\left. {\quad \quad \quad \quad \quad \quad \quad \quad \quad \quad  + {\Delta _p}{e^{ - 2{\gamma _0}\left( {d - z} \right)}} + 2\Delta _p^2{e^{ - 2{\gamma _0}d}}} \right)}
\end{array}}\\
{ + \left. {\left[ {\frac{{{{(\omega /c)}^2}}}{{1 - \Delta _s^2{e^{ - 2{\gamma _0}d}}}}} \right]\left( {{\Delta _s}{e^{ - 2{\gamma _0}z}} + {\Delta _s}{e^{ - 2{\gamma _0}\left( {d - z} \right)}} - 2\Delta _s^2{e^{ - 2{\gamma _0}d}}} \right)} \right\}}\\
{\begin{array}{*{20}{l}}
{T_{zz}^1 =  - \int_0^\infty  {\frac{{dq{q^3}}}{{{\gamma _0}}}} \frac{{{J_0}(q\rho ){\Delta _p}}}{{1 - \Delta _p^2{e^{ - 2{\gamma _0}d}}}}}\\
{\quad \quad \quad \quad  \times \left( {{\Delta _p}{e^{ - 2{\gamma _0}z}} + {\Delta _p}{e^{ - 2{\gamma _0}\left( {d - z} \right)}} - 2\Delta _p^2{e^{ - 2{\gamma _0}d}}} \right).}
\end{array}}
\end{array}
\label{Eqny}
\end{equation}
Here $J_0(q \rho)$ is a Bessel function of the first kind,
$\gamma_i=\sqrt{q^2+\epsilon_i \omega^2/c^2}$ 
and
\begin{eqnarray}
{\Delta _p} = \frac{{{\gamma _1} - {\gamma _0}{\varepsilon _1}}}{{{\gamma _1} + {\gamma _0}{\varepsilon _1}}};\quad {\Delta _s} = \frac{{{\gamma _1} - {\gamma _0}}}{{{\gamma _1} + {\gamma _0}}}.
\end{eqnarray}

The resonance energy in narrow channels can be calculated if the dielectric function, $\epsilon_1(i \omega)$, of the wall materials is known. In the present work we need the dielectric functions for silica (SiO$_2$), zinc oxide (ZnO), and gold (Au). The complex dielectric functions of the oxides were therefore determined employing a first-principles approach within the GW method based on the density functional theory\,\cite{BostPRB2012}. The dielectric function of gold was determined from optical data\,\cite{Sernelius}.

In the case of two identical atoms the above resonance condition can be separated in one antisymmetric and one symmetric part. Since the excited symmetric state has a much shorter life time than the antisymmetric state the system can be trapped in an excited antisymmetric state\,\cite{Bostrom1}.  The resonance interaction energy of this antisymmetric state is,
\begin{equation}
\label{Eq5}
U(\rho)= \hbar [\omega_{r} (\rho)-\omega_{r} (\infty)].
\end{equation}
Since the relevant solution of  Eq.\,(\ref{Eq3}) really is the pole of the antisymmetric part of the underlying Green function we can in a standard way\,\cite{Sernelius} deform a contour of integration around this pole to obtain a both simple and exact expression for the resonance interaction energy,
\begin{eqnarray}
U(\rho) = (\hbar/ \pi) \int_0^\infty d \xi \ln[1+\alpha(1|i \xi) T(\rho|i \xi)].
\label{Eq6}
\end{eqnarray}

\begin{figure}
\includegraphics[width=8.0cm]{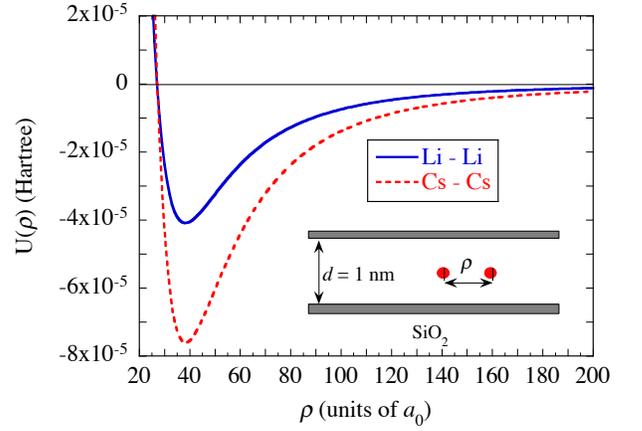}
\caption{(Color online) Full interaction energy between two symmetrically excited lithium atoms and between two symmetrically excited cesium atoms  at the center of a 1 nm planar channel in silica.}
\label{figu2}
\end{figure}

\begin{figure}
\includegraphics[width=7.2cm]{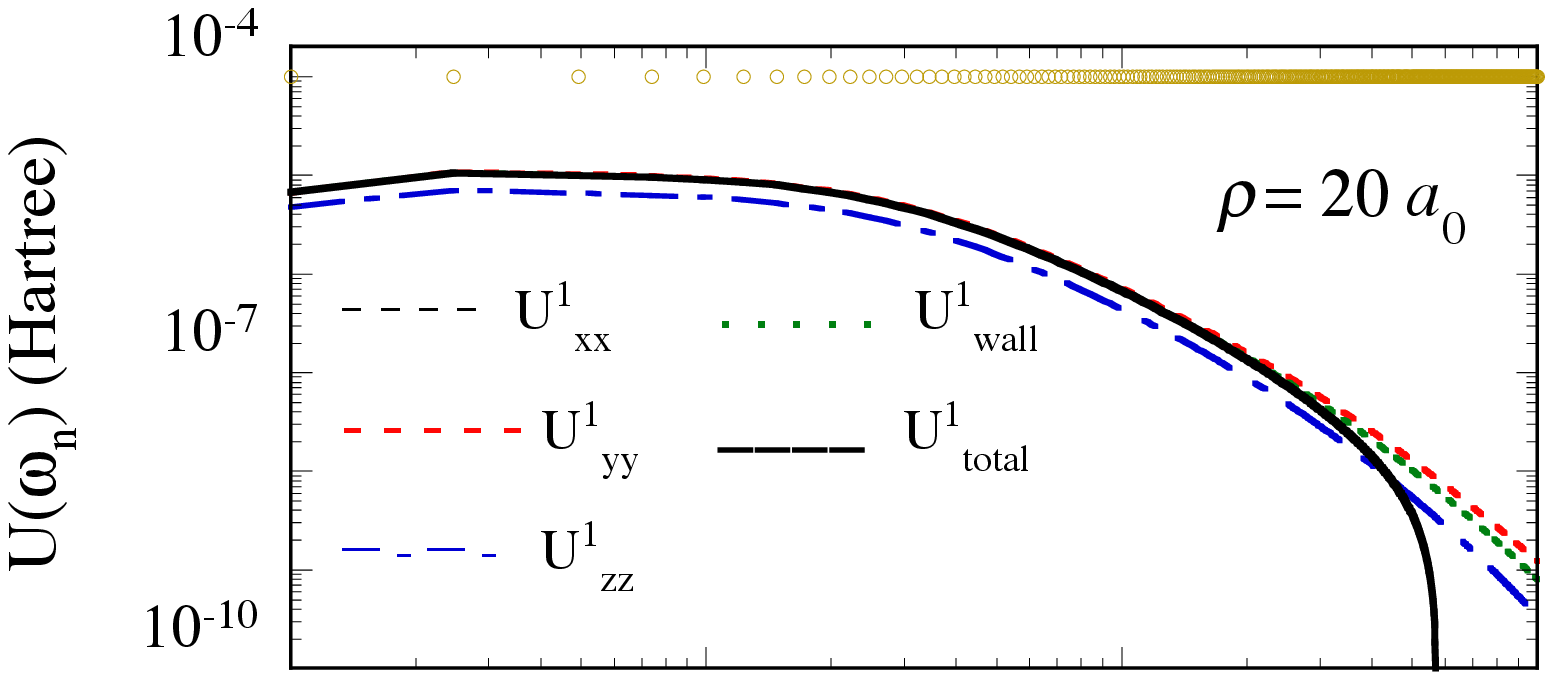}
\includegraphics[width=7.2cm]{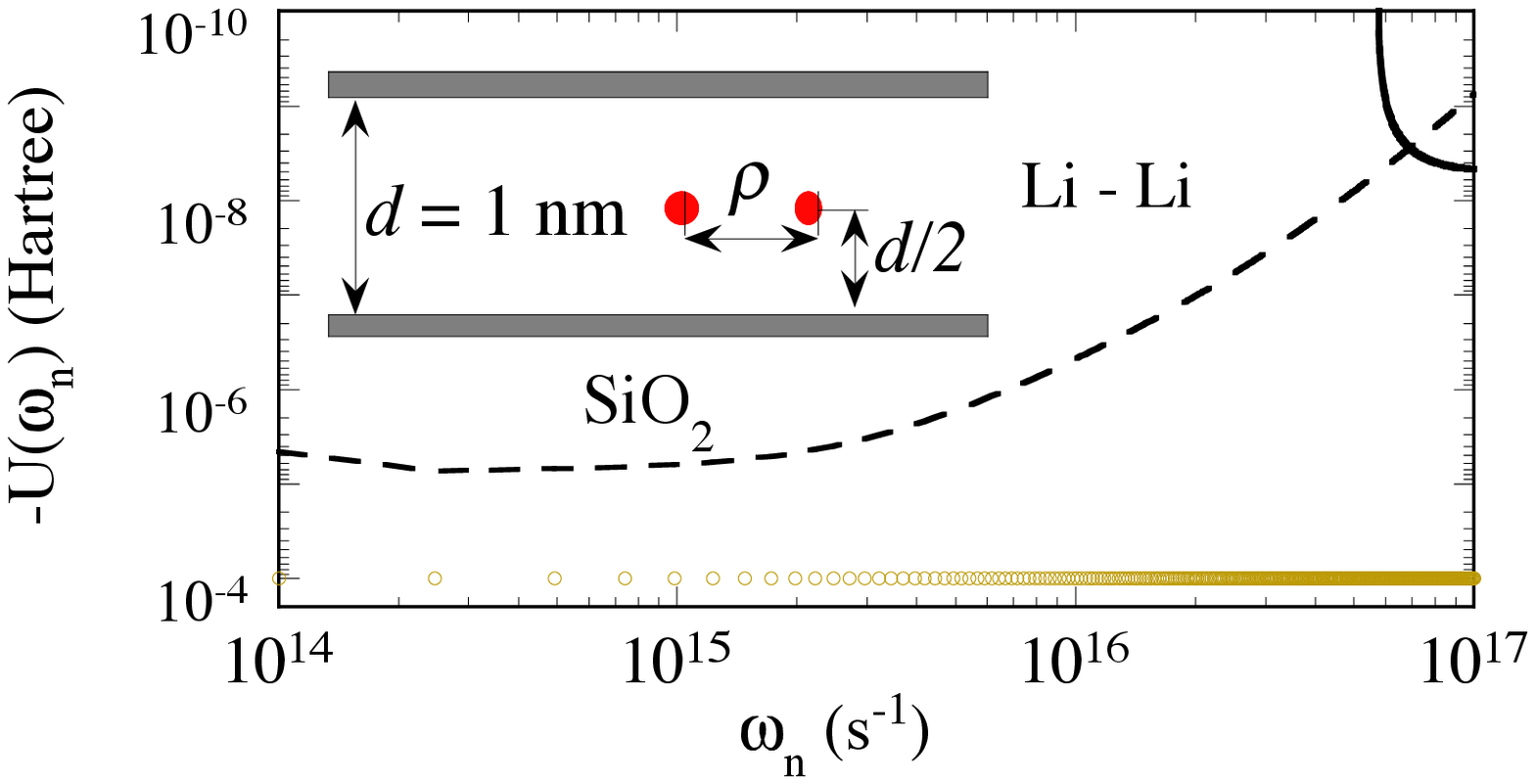}
\caption{(Color online)  Contribution from  each term $n$ for the system shown in Fig.\,\ref{figu2}   for a case when the total interaction energy is repulsive ($\rho=20 a_0$). The frequency decomposition of the interaction energy from $T_{xx}^1$,  $T_{yy}^1$,  $T_{zz}^1$  contributions and from the sum of these surface corrections, $U_{xx}^1$,  $U_{yy}^1$,  $U_{zz}^1$ and $U_{wall}^1 = U_{xx}^1 + U_{yy}^1 + U_{zz}^1$, respectively, is shown. Also shown is the total interaction energy contribution, $U_{total}^1$, including both surface corrections and the free space result from different frequencies. The discrete frequencies are shown at the top and at the bottom as circles. Note that for illustrative purposes the zero frequency contribution has been moved to the vertical axis.}
\label{figu3}
\end{figure}

To account for the temperature ($T$) dependence we simply replace the integration over imaginary frequencies by a summation over discrete Matsubara frequencies\,\cite{Sernelius,Mahanty},
\begin{equation}
\frac{\hbar }{{2\pi }} \int_0^\infty d \xi \rightarrow k_B T
\sum_{n=0}^{\infty}{'},\quad \xi_n=2 \pi k_B T n/\hbar, 
\label{Eq34}
\end{equation}
where $k_B$ is the Boltzmann constant and the prime indicates that the $n=0$ term should be divided by 2. 
Focusing first on what happens in free space we find that the correct leading term, at large separations, when the modes in the ($\pm$;$x$) branch are excited, is
\begin{equation}
U(\rho,T)\simeq \pm \frac{{2{k_B}T}}{{{\rho ^3}}}\sum_{n=0}^{\infty}{'} \alpha(i \xi_n) e^{-x n} [1+x n+x^2 n^2],
\label{Eq35} 
\end{equation}
where $x=2 \pi k_B T \rho/(\hbar c)$. We can replace the polarizability with $\alpha (0)$ at intermediate and large separations. Within this approximation the resonance free energy is,
\begin{equation}
\begin{array}{l}
U(\rho ,T) \simeq  \pm \frac{{2{k_B}T\alpha (0)}}{{2{\rho ^3}{{({e^x} - 1)}^3}}}\\
 \quad \times [1 + {e^{3x}} - {e^x}(1 + 2x - 2{x^2}) + {e^{2x}}( - 1 + 2x + 2{x^2})].
\end{array}
\label{Eq35b} 
\end{equation}
 For intermediate values of $x$ (see the middle portion of the full result in Fig.\,\ref{figu1}) this free energy of resonance interaction varies as $1/\rho^4$:
\begin{equation}
U(\rho) \simeq \pm 4 \hbar c \alpha (0)/(\pi \rho^4).
\label{AEq32}
\end{equation} 
This represents the dominating term in the interaction energy for oscillators in an excited configuration at zero temperature.   However, for any finite temperature at sufficiently large distances the long-range interaction is dominated by the $n=0$ term. This term is here
\begin{equation}
U(\rho ,T)_{n=0}=\pm k_B T \alpha(0)/\rho^3.
\label{Eq36}
\end{equation}
 This is the correct asymptotic long-range resonance interaction at any finite temperature (see the rightmost portion of the full result in Fig.\,\ref{figu1}.) 

 As a first illustration of the full result and of several limiting expressions we present,  in Fig.\,\ref{figu1}, the results for two lithium atoms in free space when the modes in the ($\pm$;$x$) branch are excited.

 An interesting case studied in Fig.\,\ref{figu2} is when the lithium or cesium atoms are excited symmetrically (i.e. when they are averaged over all possible orientations into a symmetric final state). The total resonance interaction produces short range repulsion and long range attraction creating the possibility to have bound states. In other words very large molecules can be formed inside pores. As we show in the figure the attractive potential well is about twice as deep when lithium atoms are replaced by cesium atoms.

\begin{figure}
\includegraphics[width=7.2cm]{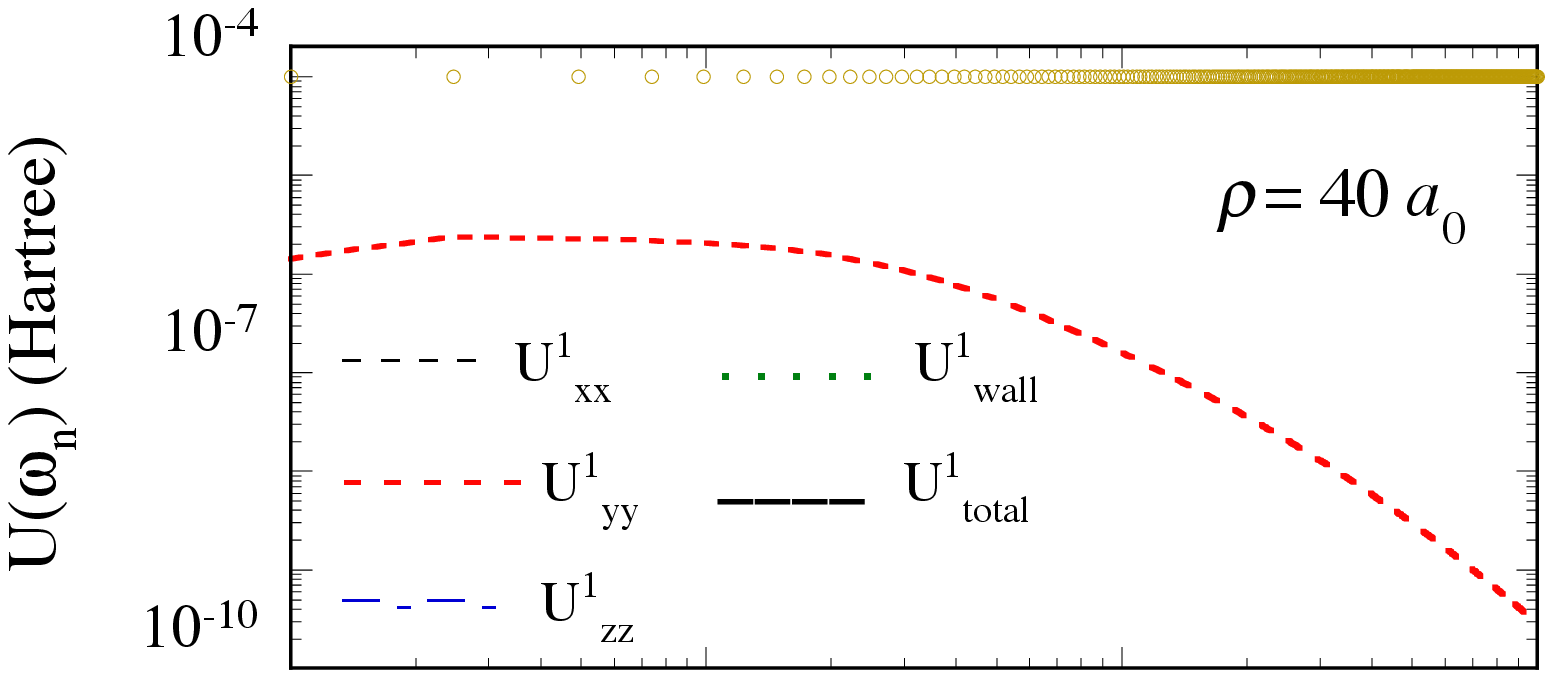}
\includegraphics[width=7.2cm]{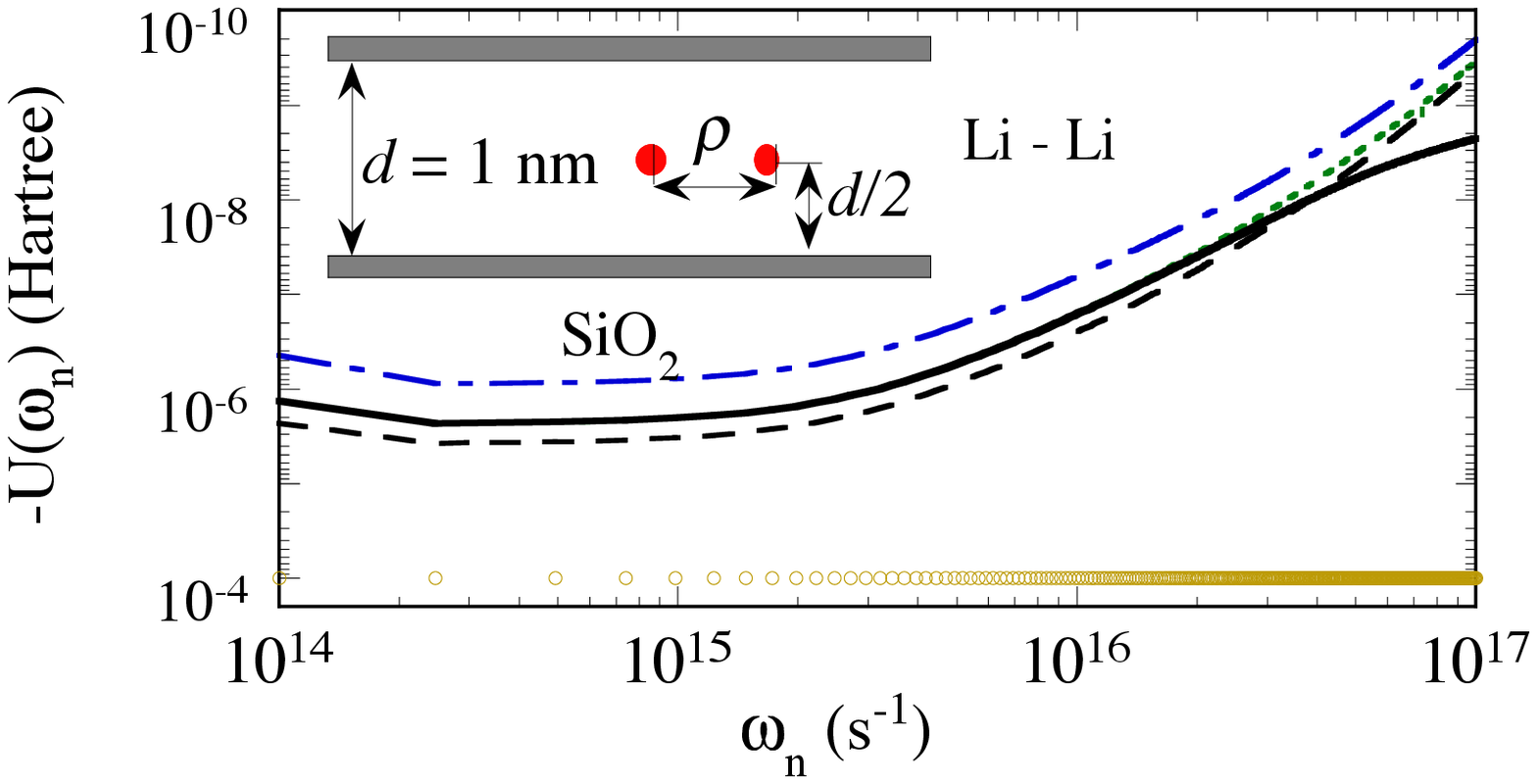}
\caption{(Color online) Same as  Fig.\,\ref{figu3} but for a larger separation ($\rho=40 a_0$) between the lithium atoms so that the total interaction energy is attractive. }
\label{figu4}
\end{figure} 

\begin{figure}
\includegraphics[width=8.0cm]{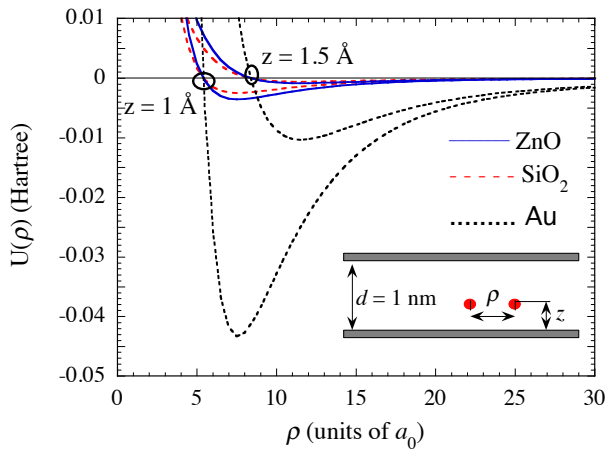}
\caption{(Color online) Full interaction energy between two symmetrically excited lithium atoms  in a 1 nm planar channel in silica (dashed curves), in zinc oxide (solid curves) and in gold (dotted curves). Here the atoms are positioned at the distance $z$ from one of the walls. }
\label{figu5}
\end{figure}

We show in Figs.\,\ref{figu3} and\,\ref{figu4}  how different modes contribute to a total repulsive short range interaction and a long range attraction. 
In the free space case attractive and repulsive interactions from the different branches cancel out in the non-retarded case.  However, a different scenario occurs when the modes are confined. The different branches are squeezed out differently when the atoms are within a planar cavity. The $x$ branch is directed between the atoms while the $z$ branch is perpendicular to the plates (the $y$ branch is perpendicular to the $z$ and  $x$ branches).
The short range repulsion between two atoms in the cavity comes from the repulsive $y$ and $z$ branches being larger than the attractive  $x$ branch. As the distance between the atoms increases only the $y$ branch gives repulsion and the total interaction is attractive.

\begin{figure}
\includegraphics[width=8.0cm]{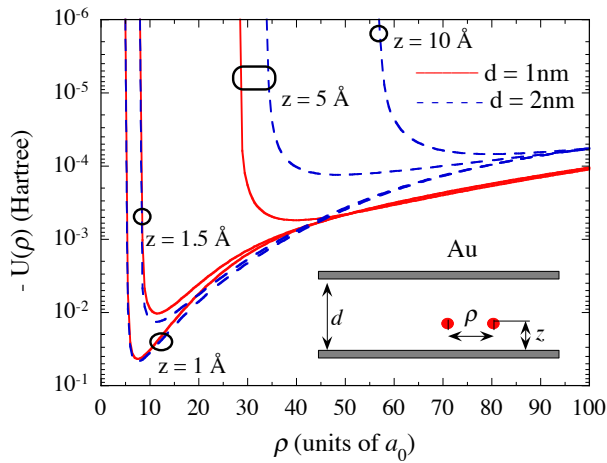}
\caption{(Color online)  Minus the full interaction energy between a pair of lithium atoms in a symmetrically excited final state between two gold surfaces distances $d=1nm$ and $d=2nm$ apart.}
\label{figu6}
\end{figure}

The depth of the attractive potential well is much larger if the atoms move away from the center of the channel and approach one of the boundaries. This is illustrated in Fig.\,\ref{figu5}.  The attractive potential well is much larger for atoms in between gold surfaces as compared to between silica or zinc oxide surfaces. In  Fig.\,\ref{figu6} we show that the binding energy increases as the atoms move closer to one of the gold surfaces. The size of the molecule is largest at the center of the cavity and increases with channel width. However, for too large cavities the binding energy in the center will be too small to effectively bind the atoms together. Close to surfaces, independent of channel width, there will be sufficient attractive interaction to bind atoms together.

The focus of this paper is on resonance interactions when excited atom pairs are inside nanochannels,  e.g.  inside porous silica or between gold walls. Resonance interactions can be suppressed inside channels when the excitation is not symmetric. For a symmetric excitation in free space the resonance interaction energy averages to zero in the non-retarded limit. A very different scenario evolves inside a planar nanochannel. Here there can be short range repulsion followed by long range attraction creating a possibility to have a bound state. This means that there can be very large molecules with a size that depends on the channel width. The binding energy is dramatically enhanced close to one of the two surfaces. Casimir-Polder interaction between atoms and surfaces drives the pair towards the surface where the size of the molecules decreases. 
The analysis of the resonance interactions undertaken here and its application to our particular case of atoms in a nanochannel show some new features of interactions in confined geometries. Interactions in, e.g., cylinders  can be tackled straightforwardly by available semi classical techniques\,\cite{Mahanty}. Already, it is clear from the example we have studied that  confinement geometry at nanoscales can produce qualitatively new results. We have confined the study to two atoms at the same distance from the surfaces. An extension of the work would be to investigate how the effects are modified when one atom is closer to a surface than the other. One knows that the decay rate of a dipole changes in a cavity and depends on the orientation\,\cite{Agar}. The formation of these large scale molecules requires two atoms where one is in an excited state. This could be difficult to achieve in a cavity with thick walls especially if they are made of gold. One way to manage this for a metallic cavity could be to excite the atoms outside, before they enter the cavity. Another way might be to have one wall made of glass coated by a thin ITO (indium tin oxide) layer or of ZnO where the inner surface layer is heavily doped; this wall would be transparent in the visible and ultra violet parts of the spectrum and behave as a metal for lower frequencies.  Then the atoms could be excited with a laser through this prepared wall. To describe this an extended Green tensor for stratified media has to be used. This is available in the literature\,\cite{Paulus}.

The observed effects could possibly be studied and verified using gas chromatography with new filter types.

Finally we may speculate in some possible applications of the observed phenomena: design of more effective lubricants; better selection of molecules for "electronic noses"; more effective surface reactions for ${H_2}$ generation; pressure sensitive filtering of gas flow in nanomaterials; new materials for gas chromatography.

 MB acknowledges support from an European Science Foundation exchange grant within the activity "New Trends and Applications of the Casimir Effect", through the network CASIMIR. B.E.S. acknowledges financial support from VR (Contract No. 70529001).


\begin{thebibliography}{10}
%
%
\bibitem{Blum} Z. Blum and S.T. Hyde,   and B. W. Ninham, J. Phys. Chem.{\bf 97}, 661 (1993).
\bibitem{Hopmeier} M. Hopmeier, W. Guss, M. Deussen, E. O. G\"{o}bel, and R. F. Mahrt, Phys. Rev. Lett. {\bf 82}, 4118 (1999).
\bibitem{Agar} G. S. Agarwal, and S. Dutta Gupta, Phys. Rev. A {\bf 57}, 667 (1998).
\bibitem{Bostrom1} M. Bostr\"{o}m, J.J.   Longdell, D.J.  Mitchell, and B.W.  Ninham, Eur. Phys. J. D {\bf 22}, 47 (2003). 
\bibitem{BostromEPL} M. Bostr{\"o}m, J.J. Longdell, and B.W.  Ninham,  Eur. Phys. Lett. {\bf 59},  21 (2002).
\bibitem{Hartman} R. L. Hartman and P. T. Leung, Phys. Rev. B {\bf 64}, 193308 (2001).
\bibitem{Sherkunov} Y. Sherkunov,  Phys. Rev. A {\bf  75},  012705 (2007). 
\bibitem{Stephen}  M.J. Stephen,  J. Chem. Phys. {\bf 40},  669 (1964).
\bibitem{McLachlan} A.D. McLachlan, Molec. Phys. {\bf 8}, 409 (1964).
\bibitem{Boser} Bo E. Sernelius, Int. J. of Modern Phys. A (IJMPA) {\bf 24}, 1804 (2009).
\bibitem{Marinescu} M. Marinescu and L. You, Phys. Rev. A {\bf 59},  1936 1999.
\bibitem{Rahmani} A. Rahmani, P.C.  Chaumet, F.   de Fornel, and C. Girard,  Phys. Rev. A {\bf 56},  3245 (1997).
\bibitem{BostPRB2012}  M. Bostr\"{o}m, Simen \AA. Ellingsen, I. Brevik, M. Dou,  C. Persson, and B. E. Sernelius, (unpublished).
\bibitem{Sernelius} Bo E. Sernelius,  {\it Surface Modes in Physics}, Wiley, Berlin, 2001. 
\bibitem{Mahanty} J. Mahanty, and B. W.  Ninham, {\it Dispersion Forces}, Academic Press: London and New York, 1976.
\bibitem{Paulus} M. Paulus, P. Gay-Balmaz, and O. J. F. Martin, Phys. Rev. E {\bf 62}, 5797 (2000).

\end{thebibliography}
\end{document}